\documentclass[twocolumn,prl,superscriptaddress,amsmath,amssymb,letterpaper]{revtex4}
\usepackage{graphicx}
\usepackage{bm}

\def\beq{\begin{equation}}
\def\eeq{\end{equation}}
\def\beqn{\begin{eqnarray}}
\def\eeqn{\end{eqnarray}}

\begin{document}

\title{Optical coupling to spin waves in the cycloidal multiferroic BiFeO$_3$}
\author{Rogerio de Sousa}
\affiliation{Department of Physics, University of California,
Berkeley, CA 94720}
\author{Joel E. Moore}
\affiliation{Department of Physics, University of California,
Berkeley, CA 94720} \affiliation{Materials Sciences Division,
Lawrence Berkeley National Laboratory, Berkeley, CA 94720}
\date{\today}

\begin{abstract}
  The magnon and optical phonon spectrum of an incommensurate
  multiferroic such as BiFeO$_3$ is considered in the framework of a
  phenomenological Landau theory. The resulting spin wave spectrum is
  quite distinct from commensurate substances due to soft mode
  anisotropy and magnon zone folding. The former allows electrical
  control of spin wave propagation via reorientation of the
  spontaneous ferroelectric moment. The latter gives rise to multiple
  magneto-dielectric resonances due to the coupling of optical phonons
  at zero wavevector to magnons at integer multiples of the cycloid
  wavevector.  These results show that the optical response of a
  multiferroic reveals much more about its magnetic excitations than
  previously anticipated on the basis of simpler models.
\end{abstract}
%
\maketitle
%

The coupling between different types of correlated electron order in
``multiferroic'' materials possessing simultaneous magnetism and
ferroelectricity gives rise to several interesting phenomena
\cite{wang03,kimura03,pimenov06,sushkov07}.  This coupling leads to
rather complex broken-symmetry phases~\cite{harris06} and excitation
spectra.  Recent examples are the magnetically induced
ferroelectricity in TbMnO$_3$\cite{kimura03} and the switching of the
magnetic state by an electrical probe observed in
BiFeO$_3$~\cite{zhao06}.  Crystals of these two materials show
incommensurate magnetic order: in TbMnO$_3$, the incommensurate order
established at 41 K is believed to generate the spontaneous
polarization observed at 28 K, while in BiFeO$_3$, the cycloidal
antiferromagnetic order is established at 650 K, well below the
ferroelectric transition at 1120 K.  The cycloid disappears in
a strong (18 T) magnetic field~\cite{ruette04} or in thin films~\cite{bai05}.

This paper studies the excitations and electromagnetic response in a
cycloidal multiferroic such as bulk BiFeO$_3$.  The combination of
incommensurate order and magnetoelectric coupling is found to give a
strong coupling between a long-wavelength electric field and spin
waves at multiples of the cycloid wavevector.  This leads to a series
of resonances in the dielectric constant at integer and irrational
multiples of a fundamental frequency determined by the magnon energy
at the cycloid wavevector.  The lowest-lying resonances created by
this effect should be visible in standard transmission and
reflectivity measurements at room temperature in the far infrared
frequency range (below the lowest optical phonon at 2~THz).  Recent
Raman and optical reflectivity spectroscopy studies of bulk BiFeO$_3$
\cite{haumont06} focused on the optical phonon resonances, without any
interpretation of the far IR region.  The optical phonon spectra seem
to be well understood from first principles calculations
\cite{hermet07}.  Nevertheless, the latter completely ignores the
underlying cycloidal magnetic structure and its magnetoelectric
character. Our work reveals a rich sub-phonon dielectric response not
yet explored by optical experiments or first principle calculations.
Aside from allowing a direct observation of the basic physics of
multiferroic coupling in BiFeO$_3$ and related materials, this effect
could be used in devices based on electronic excitation of spin waves
\cite{khitun05}, or the development of fast magnetic probes for
magnetic domain switching.

It has been known for many years that magnetoelectric coupling in
uniformly ordered materials mixes spin waves (magnons) and
polarization waves (optical phonons) \cite{baryakhtar69,tilley82}.
This gives rise to low-frequency magneto-optical resonances in the
dielectric susceptibility, the so-called electromagnon excitations,
that may be visible in infrared experiments
\cite{pimenov06,sushkov07}.  However, the resonance frequencies are
primarily determined by the zero-wavevector magnon frequencies.  The
mixing with finite-wavevector magnons found here results specifically
from the {\it incommensurate} magnetic order; while such
incommensurate order complicates the calculation of excitation spectra
because a finite-wavevector analysis is now necessary, incommensurate
order is a common feature of many of the most studied multiferroic
materials.  Although recent theoretical work has clarified the
relationship between ferroelectricity and spiral ferromagnetic order
\cite{mostovoy06,katsura05}, the characteristic features of
electromagnon excitations in incommensurate magnets seem not to have
been studied before.

This paper assumes that the ground state consists of a uniform
polarization and an incommensurate magnetic structure, as in many
important multiferroic materials; an interesting effect when the
polarization and magnetization are both periodic and commensurate with
each other has recently been discussed \cite{betouras07}.

Our calculation is based on a dynamical Ginzburg-Landau treatment of
the coupled ferromagnetic, antiferromagnetic, and polarization orders,
in a cycloidal ground state and an applied AC electrical field.  It is
assumed that the system is far below its critical temperature so that
thermal fluctuations can be ignored in the dynamics.  The long-period
incommensurate cycloid is produced in our model by a Lifshitz term in
the effective Ginzburg-Landau free energy; the existence of such a
term was previously
argued~\cite{sparavigna94,sosnowska82,zalesski00,ruette04} as the
unique symmetry-allowed explanation for the observed order.  The model
free energy is
\begin{eqnarray}
F &=& {G L^4 \over 4} + {A L^2 \over 2} + {c \sum_i
  (\nabla L_i)^2 \over 2} \cr
&& -\alpha \bm{P} \cdot \left[{\bf L} (\nabla \cdot {\bf L}) + {\bf L} \times
(\nabla \times {\bf L})\right] -\bm{P}\cdot \bm{E} \nonumber\\
&&+{r M^2\over 2}+ {a {P_z}^{2}\over 2}+{u{P_z}^{4}\over 4} + {a_\perp ({P_x}^2+{P_y}^2) \over 2} .
\end{eqnarray} 
Here $L=|\bm{M}_1-\bm{M}_2|$ is a N\'{e}el vector describing the
staggered sublattice magnetization, $M=|\bm{M}_1+\bm{M}_2|$ is the
total magnetization of the material, and $P_z$ is the magnitude of the
ferroelectric polarization along ${\bf \hat z}$ [the cubic (111) and
equivalent directions in BiFeO$_3$].  Many possible terms have been
omitted from the free energy as absent or unimportant in BiFeO$_3$.
Note that in the absence of the Lifshitz term, the AFM and FE orders
are decoupled, and the ground state is a commensurate, isotropic
Heisenberg antiferromagnet with $|{\bf L}_0|^{2}= {- A \over G}$ and
an easy-axis ferroelectric with uniform polarization given by
$\bm{P}_0=P_0 {\bf \hat{z}}$, $P_{0}^{2}= -a/u$.  Clearly $A < 0$ and
$r>0$ for an antiferromagnetic ground state, while $a<0$ and
$a_{\perp}>0$ for a ferroelectric ground state.  
%
%

\begin{figure}
  \includegraphics[width=3in]{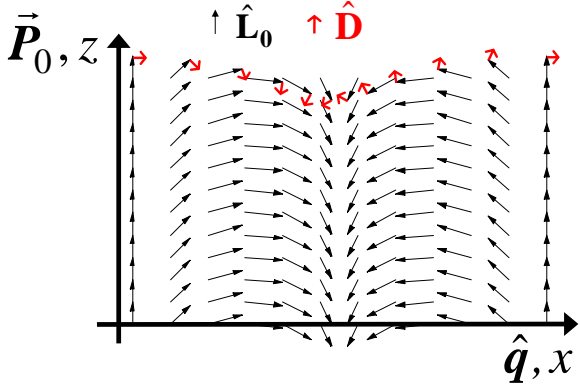}
\caption{(color online). Antiferromagnetic lattice with harmonic cycloid order. 
  The cycloid plane is pinned by the polarization $\bm{P}_0$, with
  arbitrary propagation direction ${\bf \hat{q}}$ in the plane
  perpendicular to $\bm{P}_0$. We also depict the tangent vector ${\bf
    \hat{D}}$ related to phase fluctuations of the cycloid.}
  \label{fig1} \end{figure} 

The Lifshitz term induces an incommensurate cycloidal
order in which the antiferromagnetic moment rotates in an arbitrary
plane including ${\bf \hat z}$.  It also increases the magnitude of
$L_0$ and $P_0$.  Our reference ground state is a
cycloid with 
\begin{equation} 
{\bf L}_0(x) = L_0 \left[\cos(q x) {\bf \hat z}
  + \sin(qx) {\bf \hat x}\right].  
\label{l0}
\end{equation} 
Here the cycloid direction has been chosen along ${\bf \hat x}$.  The
pitch is $q = \alpha P_0 / c$ and the magnitudes of antiferromagnetic
and polar order are 
$L_{0}^{2} = (-A+\alpha^2 P_{0}^{2}/c)/G$,
$P_{0}^{2}=(-a+\alpha^2 L_{0}^{2}/c)/u$.  A check on the above is that
there are two zero-energy symmetry actions for fixed ${\bf P}$:
changing the cycloid phase and rotating the cycloid direction
in the $xy$ plane.  The displacement for the phase change from the
transformation $qx\rightarrow qx + \phi/L_0$ is
\begin{equation}
\delta \bm{L}=\phi \left[\cos(qx){\bf \hat{x}} -\sin(qx){\bf \hat{z}}\right].
\end{equation}
We can use the shorthand ${\bf \hat D}(x)$ for this spin vector
direction transverse to the local ground-state spin direction, and
write $\delta {\bf L} = \phi {\bf \hat D}(x)$ for the phase change.
The displacement for the cycloid direction change due to an
infinitesimal rotation $q{\bf \hat{x}}\rightarrow q({\bf \hat{x}}+\eta
{\bf \hat{y}})$ is 
$\delta {\bf L} = \eta L_0 q y {\bf
  \hat D}(x) + \eta L_0 \sin (q x) {\bf \hat y}$. This cycloid
direction symmetry is similar to the one found in smectic liquid
crystals in that it requires large displacements relative to the
original state at some point in the crystal.  

The linearized equations of motion for the ferroelectric and
antiferromagnetic order parameters are
\begin{subequations}
\begin{eqnarray}
\frac{\partial^{2} \bm{L}}{\partial t^{2}}
&=&-r(\gamma L_0)^2 \left[\frac{\delta F}{\delta \bm{L}}-\left(
{\bf \hat{L}}_{0}\cdot 
\frac{\delta F}{\delta \bm{L}}
\right)
{\bf\hat{L}}_{0}
\right],\label{ll}\\
\frac{\partial^{2} \bm{P}}{\partial t^{2}}&=&-f \frac{\delta F}{\delta \bm{P}}.\label{debye}
\end{eqnarray}
\end{subequations}
Here $\gamma$ is a gyromagnetic ratio with dimensions of (sG)$^{-1}$,
while $f$ has dimensions of s$^{-2}$ and plays a similar role in the
ferroelectric equation of motion. In the above polarization equation, we have
ignored damping and the possible Poisson-bracket term between ${\bf P}$ and
${\bf M}$~\cite{bakerjarvis01}. The coupled spin and polarization equations are solved in terms of the parametrization
\begin{subequations}
\begin{eqnarray}
\delta \bm{L}&=& \phi(\bm{r}) \textrm{e}^{-i\omega t}{\bf \hat{D}}(x) 
+\psi(\bm{r}) \textrm{e}^{-i\omega t} {\bf \hat{y}},\\
\delta \bm{P}&=& \delta \bm{p}(\bm{r}) \textrm{e}^{-i\omega t}.
\end{eqnarray}
\end{subequations}
The field $\phi(\bm{r})$ denotes phase fluctuations of the cycloid
ground state, while $\psi(\bm{r})$ refers to spin fluctuations out of
the cycloid ($xz$) plane. The field
$\delta\bm{p}(\bm{r})=\bm{P}-\bm{P}_0$ denotes optical phonon
fluctuations related to longitudinal and transverse vibrations of the
ferroelectric moment $\bm{P}$ \footnote{While there are many optical
  phonons in a ferroelectric, this field $\bm{P}$ should be thought of
  as describing the particular polar phonon that goes soft at the
  ferroelectric transition.}.

\begin{widetext}
Calculating $\delta F/\delta \bm{L}$ to linear order in the
displacements, and inserting into Eq.~(\ref{ll}) leads to
\begin{subequations}
\begin{eqnarray}
\left[\omega'^2+c\nabla^2\right]\phi&=&2cq\sin(qx)(\partial_y \psi)
-\alpha L_0 \left(\nabla\times \delta\bm{p}\right)
\cdot{\bf \hat{y}},
\label{phieqn}\\
\left[\omega'^2-cq^2+c\nabla^2\right]\psi&=&
-2cq\sin(qx) (\partial_y\phi)
-\alpha L_0 \left[
2q\cos(qx)(\delta p_y) -\left(\nabla\times \delta\bm{p}\right)\cdot {\bf \hat{D}}
\right],\label{psieqn}
\end{eqnarray}
with $\omega'=\omega /(\gamma L_0 \sqrt{r})$ defining the frequency in
units of $\sim 10^{12}$~rad/s.  The equation of motion for the
polarization becomes
\begin{equation}
\xi \omega'^2 (\delta\bm{p})=\frac{\delta F}{\delta \bm{P}}=
a_{\parallel}(\delta\bm{p}_z) +a_{\perp}(\delta\bm{p}_{\perp})
-\alpha L_0 \left[
q\cos(qx)(\psi) {\bf \hat{y}} +(\bm{\nabla} \phi) \times {\bf \hat{y}}-(\bm{\nabla}\psi) \times {\bf \hat{D}}
\right]-\bm{E},\label{dpeqn}
\end{equation}
\end{subequations}
where we define $\xi = r(\gamma L_0)^2/f$  ($\sim 10$ in BiFeO$_3$),
$a_{\parallel}=-2a+3\alpha^2L_{0}^{2}/c$, and the AC electric field is
$\bm{E}\textrm{e}^{-i\omega t}$.
\end{widetext}

Some intuition for Eqs.~(\ref{phieqn})-(\ref{dpeqn}) may be gained by
considering their invariance
under rotations of the cycloid in the $xy$ plane. Consider the
$\omega=0$ symmetry operation $\phi\rightarrow \eta q y $, $\psi
\rightarrow \eta \sin(qx)$, with $\delta\bm{p}$ left unchanged.  This
creates no terms in both sides of Eq.~(\ref{phieqn}). In
Eq.~(\ref{psieqn}), the left hand side becomes $-2cq^2\eta \sin(qx)$,
which cancels the term generated in the right hand side due to
the $y$-dependence in $\phi$.  For the square brackets in
Eq.~(\ref{dpeqn}), the first term adds $q\cos(qx)[\eta\sin(qx)]{\bf
  \hat{y}}$, with the third term $-\bm{\nabla}[\eta\sin(qx)]\times{\bf
  \hat{D}}$ giving the desired cancellation.

In the commensurate limit, $\alpha,q \rightarrow 0$ and we recover two
transverse AFM fluctuation modes with $\omega'^2=ck^2$.  There are two
dispersionless polarization wave modes transverse to $\bm{P}_0$ with
frequency $\omega'^2=a_{\perp}/\xi$, and one longitudinal mode with
frequency $\omega'^2=a_{\parallel}/\xi$.  The combination orthogonal
to $\phi{\bf \hat{D}}$, and $\psi {\bf \hat{y}}$, $S = {\bf \hat{L}}_0
\cdot {\delta \bm{L}}$, is constant in time in the equation of motion
above but decays dissipatively if non-Poisson bracket terms are kept.
It is not expected to modify the above propagating modes.  For $k
\rightarrow 0$ in the $\phi$ mode, we obtain a symmetry (the phase
shift) which should be nondissipative.  For $k \rightarrow 0$ in the
$\psi$ mode, there should be some finite dissipation.

We solve the system Eqs.~(\ref{phieqn})-(\ref{dpeqn}) using the ansatz
\begin{equation}
[\phi(\bm{r}),\psi(\bm{r}),\delta\bm{p}(\bm{r})]= 
\sum_{n} [\phi_n,\psi_n,(\bm{p})_n] \textrm{e}^{inqx}\textrm{e}^{i\bm{k}\cdot \rm{r}},
\end{equation}
with $n$ an integer running from $-\infty$ to $\infty$ and the
restriction $|k_x|<q/2$. The complex numbers $\phi_n$, $\psi_n$, $(p_i)_n$ are
independent of $\bm{r}$ but have a $\bm{k}$ dependence.
Substitution of this ansatz into Eqs.~(\ref{phieqn})-(\ref{dpeqn}) yields a
linear system of equations whose characteristic polynomial determines
the resonance frequencies. 

Consider the unmixed spin waves in the limit
$(\bm{p})_n\rightarrow 0$. Substitution of the ansatz into
Eqs.~(\ref{phieqn}),~(\ref{psieqn}) gives
\begin{subequations}
\begin{eqnarray}
\left[ \omega'^2-c\tilde{k}^2\right]& \phi_n & - c k_y\left(\psi_{n-1}-\psi_{n+1}\right)=0,
\label{phin}\\
\left[\omega'^2 -c(\tilde{k}^2+q^2)\right] &\psi_n & + c k_y\left(\phi_{n-1}-\phi_{n+1}\right)=0,
\label{psin}
\end{eqnarray}
\end{subequations}
with $\tilde{k}^2=(k_x+nq)^2+k_{y}^{2}+k_{z}^{2}$.  We see that modes
propagating along the cycloid plane (with $k_y=0$) are simple plane
waves with cyclon ($\phi$) and out of plane ($\psi$) dispersions given
by $\omega'^2=c\tilde{k}^2$ and $\omega'^2=c(\tilde{k}^2+q^2)$
respectively. As expected from symmetry, the cyclon mode remains soft,
but the out of plane mode $\psi$ acquires a gap due to the pinning of
the cycloid plane by the ferroelectric moment. Note that the $\psi_0$
gap equals the cyclon energy at $n=\pm 1$. At $\bm{k}=0$, the AFM
resonance modes are simply $\omega'_{n}=\sqrt{c}q|n|$ and
$\omega'_{n}=\sqrt{c}q\sqrt{n^2+1}$ for the cyclon and out of plane
excitations respectively.  For BiFeO$_3$, $cq^2\sim 1$
\cite{sosnowska82,zalesski00}, and the modes are equally spaced by approximately
$10^{12}$~rad/s intervals ($0.16$~THz or $10$~K).

If the propagation vector has a small projection along the $k_y$
direction, the dispersion curves in the cycloid plane will repel each
other whenever they intersect. This leads to a series of small gaps
(anticrossings) in the propagation frequency as a function of the
reduced $k_x$, similar to the effect discussed by Bar'yakhtar in an
helical ferromagnet \cite{baryakhtar70}. Apart from these small gaps,
the high-frequency dispersion curves are nearly unaffected by the
incommensurate order.  A very different situation is found for the
low-frequency modes along the non-trivial $y$ direction. Consider for
example the lowest frequency mode $\phi_0$. Solving
Eqs.~(\ref{phin}),~(\ref{psin}) within ${\cal O}(k^{6})$ leads to
\begin{equation}
\omega'^2\approx c\left[k_{x}^{2}+k_{z}^{2}+\frac{3}{8}\frac{k_{y}^{4}}{q^2}
-\frac{k_{x}^{2}k_{y}^{2}}{q^2}\right].
\end{equation}
This soft mode dispersion is strongly anisotropic, and may be useful
for electrical control of the spin wave group velocity via
switching of the $\bm{P}_0$ direction. A similar effect is found for
phase fluctuations in smectic liquid crystals, and in an helical
ferromagnet such as MnSi \cite{kirkpatrick05}. The anisotropy is
related to the $\bm{q}$ rotation symmetry discussed above, which
forbids a $k_{y}^{2}$ term in the dispersion.

\begin{figure}
  \includegraphics[width=3in]{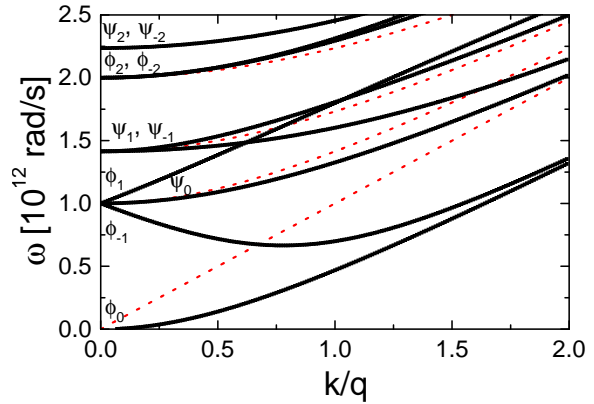} \caption{(color online). Spin
    wave spectra for a cycloidal multiferroic including one optical
    phonon mode (not shown).  The solid lines are for propagation along the
    non-trivial direction perpendicular to the cycloid plane ($y$ in
    Fig.~1). The dashed lines are the spectra for propagation along
    $z$, including zone folding.  The magnon modes at $nq$ are
    labelled by $\phi_n$ (cyclon) and $\psi_n$ (out of plane).}
  \label{fig2} \end{figure} 

A full numerical solution for propagation along $y$ is shown in
Fig.~\ref{fig2}. Note that in addition to the soft mode anisotropy
effect, $k_y\neq 0$ admixes the modes at $nq$ and $-nq$, splitting
their dispersions. Moreover, the second lowest frequency mode
(connected to $\phi_{\pm 1}$) acquires a negative group velocity.

The full solution of Eqs.~(\ref{phieqn})-(\ref{dpeqn}) with
$(\delta\bm{p})\neq 0$ shows avoided crossings between the optical phonon dispersion
and magnon branches at finite $k$. At these
anticrossings the spin wave excitations are highly mixed with the
polarization wave. However, an optical experiment only probes modes at
$\bm{k} \approx 0$ due to the large value of the velocity of light when
compared to the magnon velocity ($\sim 10^{6}$~cm/s). One can ask
whether any $n>0$ mode responds to optical excitation
with wavelength much larger than $2\pi/q\approx 600$~\AA. The answer
is affirmative as now shown by calculation of
the AC electric susceptibility.

An AC electric excitation $\bm{E}\textrm{e}^{-i\omega t}$ may excite
spin waves due to the appearence of magnetoelectric fields
such as $2q \cos{(qx)}(\delta p_y)$ 
on the right hand side
of Eq.~(\ref{psieqn}). This couples $\psi$ to an optical phonon at $k=0$.
Solving the system of equations shows that the AC electrical susceptibility tensor 
$(p_i)_0=\chi_{ij}E_{j}$ is diagonal for our model, with $\chi_{xx}=-1/(\xi
\omega'^2-a_{\perp})$, $\chi_{zz}=-1/(\xi \omega'^2-a_{\parallel})$.
Therefore the optical response with light polarized within the cycloid
plane contains no magnetic anomaly but only simple poles at the bare
optical phonon frequencies. However, if the light is polarized along
the direction perpendicular to the cycloid plane, the situation is quite different,
\begin{equation}
\chi_{yy}(\omega)=
\frac{-(\omega'^2-2cq^2)}{
(\omega'^{2}-\Omega'^{2}_{{\rm PH}})
(\omega'^{2}-\Omega'^{2}_{{\rm AFM}})}.
\end{equation}
This AC susceptibility has two poles at the following shifted phonon
and AFM ($\psi_{\pm1}$ mode) frequencies:
$\Omega'^{2}_{{\rm PH}}= a_{\perp}/\xi+2cq^2$,
$\Omega'^{2}_{{\rm AFM}}= 2cq^2-2(\alpha q L_0)^2/a_{\perp}$.

Hence we conclude that the simplest model for BiFeO$_3$ already
contains a low-frequency magneto-optical resonance. Note the prominent
importance of magnon zone folding: Even though the AC electric field
is a zero wavevector excitation, it couples to a magnon at wavevector
$\pm q$. This is a special feature of electromagnetic excitations in
an incommensurate multiferroic material. 

The magnon excitations of a cycloid with wavevector $q$ may violate
momentum conservation by a multiple of $q$ (a reciprocal lattice
vector for the cycloid). In the case of a multiferroic material with
incommensurate magnetic order, this allows the possibility of
observing additional magneto-dielectric resonances due to the coupling
of a zero-wavevector phonon to magnon modes at other multiples of
$q$.    To see this effect, consider a contribution to the free energy in the
form of an easy-plane anisotropy
$\frac{\lambda}{2}\left(\bm{P}\cdot \bm{L}\right)^2$,
which is known to add anharmonicity to the cycloid order
\cite{sparavigna94}. To first order in $\lambda$, this term adds a third
harmonic to Eq.~(\ref{l0}). This gives an additional magnetoelectric
contribution to the right side of the equation determining the
cyclon field $\phi$ [Eq.~(\ref{phieqn})]:
$\lambda P_0 L_0/2 \left[\cos{(2qx)}(\delta p_x) - \sin{(2qx)}(\delta p_z)\right]$.
As a result, the AC electric susceptibilities $\chi_{xx},\chi_{zz}$
develop poles at the magnon modes $\phi_{\pm 2}$. Similarly, the out
of plane field $\psi$ is excited by a term of the form $\cos{(3qx)}
(\delta p_y)$, showing that $\chi_{yy}$ has an additional
electromagnon pole at the mode $\psi_{\pm 3}$.  Considering additional
powers of $\lambda$ (higher harmonics of the cycloid) shows that
higher frequency resonances at $nq$ 
are present, with strength falling off as
higher powers of $\lambda$.

In conclusion, we determined the electromagnon spectra for a class of
cycloidal multiferroics including bulk BiFeO$_3$.  A simple
Landau-Ginzburg model shows that a low-frequency optical probe will
couple to magnons at multiples of the cycloid wave vector, leading to
a series of magneto-dielectric resonances at integer $n$ and
irrational $\sqrt{1+n^2}$ multiples of a fundamental cyclon frequency
($\phi_n$ and $\psi_n$ magnons respectively).  Interestingly, only one
mode ($\psi_{\pm 1}$) becomes electric dipole active due to the strong
linear magnetoelectric effect driving the harmonic cycloid order. The
remaining modes acquire a small electric dipole character due to
weaker quadratic magnetoelectric effects and anharmonicity in the
cycloid order. As a result, optical experiments such as
transmittivity, reflectivity, and Raman spectroscopy will reveal much
more about the magnetic order of multiferroics than previously
expected on the basis of homogeneous models.  For example, the
direction dependence and intensity of higher resonances is directly
related to anharmonicity in the magnetic order.  Other incommensurate
multiferroics may show similar phenomena even if the details of the
multiferroic coupling are different.

The authors acknowledge useful conversations with S. Mukerjee, J.
Orenstein, R. Ramesh, I. Souza, N. Spaldin, and A. Vishwanath. This
work was supported by WIN (RdS) and by NSF DMR-0238760 (JEM).

\end{document}